\documentclass[aps,floats,floatfix,showpacs,amssymb,prd,twocolumn,superscriptaddress,nofootinbib,nolongbibliography,reprint,preprintnumbers]{revtex4-1}

\usepackage{amssymb,amsmath,verbatim,mathtools,needspace,enumitem,etoolbox,graphicx,physics,microtype,afterpage,xspace,tabularx,lmodern,multirow,braket}
\usepackage{gensymb}
\usepackage[normalem]{ulem}
\usepackage[dvipsnames, usenames]{xcolor}
\definecolor{linkcolor}{rgb}{0.0,0.3,0.5}
\usepackage[unicode, colorlinks=true, linkcolor=linkcolor, citecolor=linkcolor, filecolor=linkcolor, urlcolor=linkcolor, linktocpage, breaklinks]{hyperref}
\usepackage[all]{hypcap}
\usepackage{subcaption}
\usepackage[T1]{fontenc}
\usepackage[utf8]{inputenc}
\usepackage[usenames,dvipsnames]{xcolor}
\definecolor{romared}{RGB}{142,0,28}
\hypersetup{colorlinks=true,citecolor=romared,linkcolor=romared,urlcolor=romared}

\newcommand{\be}{\begin{equation}}
\newcommand{\ee}{\end{equation}}

\def\be{\begin{equation}}
\def\ee{\end{equation}}
\newcommand{\beq}{\begin{eqnarray}}
\newcommand{\eeq}{\end{eqnarray}}

\usepackage{aas_macros}
\usepackage{makecell}
\usepackage{soul}
\usepackage{amssymb}

\usepackage{lipsum}

\newcolumntype{Y}{>{\centering\arraybackslash}X}

\makeatletter
\AtBeginDocument{%
  \long\def\@makecaption#1#2{%
    \vskip\abovecaptionskip
    \begingroup
      \small\rmfamily
      \@ifundefined{flushing}{\setlength\rightskip{0pt plus 1fil}}{\flushing}%
      #1\ #2\par
    \endgroup
    \vskip\belowcaptionskip}%
}
\makeatother

\begin{document}

\title{Early-time tails in Kerr black hole ringdown}

\author{Ansh Gupta}
\affiliation{Department of Physics, King’s College London,
The Strand, London WC2R 2LS, United Kingdom}

\author{Naritaka Oshita}
\affiliation{Department of Physics, Kindai University, Osaka 577-8502, Japan}
\affiliation{RIKEN iTHEMS, Wako, Saitama, 351-0198, Japan}

\preprint{RIKEN-iTHEMS-Report-26}

\begin{abstract}
Black hole ringdowns are modelled as a superposition of quasinormal modes (QNMs), although the branch cut of the Green's function produces a nonoscillatory tail that may contaminate early-time ringdown analyses.
Within linear perturbation theory, we quantify this branch-cut contribution for the scattering of localised Gaussian wave packets by a Kerr black hole with mass $M$.
Defining the onset of ringdown as the earliest time at which the QNM and branch-cut contributions reconstruct the waveform, we evaluate the branch-cut amplitude relative to the QNM amplitude.
For wave packets with a characteristic frequency $\omega_0\sim 1/M$, the branch-cut amplitude is typically of order $10\%$ of the QNM amplitude at ringdown onset with only a mild dependence on spin.
We further find that the tail waveform before convolving with a source is almost independent of spin.
The spin dependence of the relative amplitude therefore resides primarily in the excitation of the QNMs.
Our results suggest that the branch-cut contribution may affect the accuracy of black hole spectroscopy, particularly near the onset of ringdown.
\end{abstract}

\maketitle

\section{Introduction}
The dawn of gravitational-wave (GW) astronomy has ushered in an era of precision tests of gravity in the strong-field regime. The observation of GWs from binary black hole coalescences by the LIGO-Virgo-KAGRA collaboration \cite{LIGOScientific:2014pky, VIRGO:2014yos, KAGRA:2020tym, LIGOScientific:2016aoc} has established black hole ringdown, the relaxation of the remnant black hole into its stationary state, as a primary probe of this regime. Traditionally, the ringdown signal is modelled as a linear superposition of exponentially damped sinusoids, known as quasinormal modes (QNMs) \cite{Vishveshwara:1970zz,Chandrasekhar:1975zza,Kokkotas:1999bd,Berti:2009kk}, with complex frequencies which depend exclusively on the mass and spin of the remnant. Thus, black hole spectroscopy \cite{Berti:2025hly} has become an invaluable tool for characterising the properties of remnant black holes and for testing the no-hair theorem \cite{Carter:1971zc}.

This picture is, however, an approximation to the underlying linear problem rather than a complete description. Writing the response of the black hole as a Green's function and deforming the frequency integration contour into the lower half plane, as first done systematically by Leaver \cite{Leaver:1986gd} and by Nollert and Schmidt \cite{Nollert:1992ifk}, the signal separates into three distinct pieces: a prompt contribution from the high-frequency arc, encoding the direct propagation of the initial data to the observer; a sum over the poles of the Green's function, which are the QNMs; and an integral around a branch cut running down the negative imaginary frequency axis. Only the second of these is formed of damped sinusoids. The QNM sum is not a complete basis, does not converge uniformly, and cannot by itself represent the signal at arbitrarily early or late time \cite{Ching:1995tj,Ching:1995rt,Leaver:1986gd}. Therefore, ringdown modelling implicitly assumes that the neglected pieces are small over the analysis window. 

The branch-cut contribution is the origin of the familiar power-law tail, first identified by Price for a Schwarzschild background \cite{Price:1971fb,Price:1972pw}, and subsequently studied in a variety of settings \cite{Gundlach:1993tp,Ching:1995tj,Ching:1995rt}, including the Kerr spacetime \cite{Barack:1999ma,Gleiser:2007ti}. Ching \textit{et al.} \cite{Ching:1995tj} established the general link between the asymptotic behaviour of the effective potential and the resulting late-time decay law, and more recent work \cite{Rosato:2025rtr} has extended and revised that analysis to a wide class of compact objects, demonstrating that at very late times tails are universally governed by a $t^{-(2\ell+3)}$ decay law for configurations in which the effective potentials decay asymptotically as $1/r^2$ or faster. The majority of the literature is concerned with the asymptotic regime, where the tail is the dominant surviving feature of the signal. Interest in tails at intermediate times has grown sharply in the last few years, since it has been shown that they can be excited far more strongly than previously expected. For example, tails inherited from the inspiral history can be enhanced by orders of magnitude for non-circular orbits \cite{DeAmicis:2024eoy}, they are generic in dynamical spacetimes \cite{Cardoso:2024jme}, and they have now been extracted from fully nonlinear merger simulations \cite{DeAmicis:2024not,Ma:2024hzq} and analysed in eccentric binaries \cite{Islam:2024vro}. 

Whether or not tails matter for a given ringdown analysis is a question about excitation amplitudes at the times considered in the analysis. This question has become more pressing as ringdown analyses have been pushed to progressively earlier start times in order to exploit the higher signal-to-noise ratio near merger \cite{Giesler:2019uxc,Bhagwat:2017tkm,Nee:2023osy,Baibhav:2023clw,Giesler:2024hcr}. The choice of start time is already a significant systematic in QNM extraction, and any non-QNM power neglected in modelling but present in the fitting window will be absorbed into the fitted mode amplitudes, biasing the inferred remnant parameters. Therefore, if the early-time tail excitation is sufficiently large it could contaminate precise QNM extraction. On the other hand, if the tail is measurable, it offers an independent probe of the progenitor dynamics.

In this work, we quantify the early-time tail excitation for Kerr black holes as a function of the remnant spin. Using frequency-domain Green's function techniques \cite{Leaver:1986vnb, Leaver:1986gd, Andersson:1996cm}, we isolate the branch-cut contribution to the GW response of Kerr black holes and, separately, the QNM contribution, including both the prograde and retrograde (mirror) branches of the $\ell=m=2$ spectrum \cite{Dhani:2020nik}. Both components are then driven by Gaussian initial data, which allows us to control the spectral content of the perturbation. We compare the QNM and tail amplitudes at the start time which we define as the time at which the prompt response has become negligible. Our main results are that the ringdown is QNM dominated at the starting time across the range of black holes spins and initial data we consider, and this dominance increases monotonically with spin. The tail excitation has a greater relative importance at low frequencies and low spins. We also find that the shape and amplitude of the early-time tail waveform is only weakly-dependent on the black hole spin.

This paper is organised as follows. In Section \ref{sec:review}, we review the linearised problem in frequency space and show how the prompt, QNM, and tail contributions arise from the Green's function. In Section \ref{sec:response_initial}, we describe the initial data and the convolution procedure used to construct the response, and define the start time of the ringdown. Section \ref{sec:results} presents our results for the spin and frequency dependence of the QNM and tail excitations. In Section \ref{sec:disc} we summarise our results, discuss the implications for ringdown modelling, and connect our analysis to astrophysical sources. Appendix \ref{app:SN} describes the Sasaki-Nakamura formalism, Appendix~\ref{app:tail} provides the detail of computation of the branch-cut contribution in the Mano-Suzuki-Takasugi method, and Appendix \ref{app:tol} demonstrates the robustness of our conclusions with respect to the start-time criterion.
Throughout this paper, we use the units $G=c=1$.

\section{Review}\label{sec:review}
\subsection{Perturbations of the Kerr black hole}\label{sec:pert_Kerr}

We begin by reviewing the equations governing perturbations of the Kerr background.
The black hole has mass $M$ and the spin parameter $a\coloneqq J/M$, where $J$ is the angular momentum of the black hole.
The dimension-less spin parameter is $\chi \coloneqq a/M$.
In Boyer-Lindquist coordinates $(t,r,\theta,\phi)$, the line element of the Kerr black hole is
\begin{equation}
\begin{split}
    ds^2 = -&\frac{\Delta}{\Sigma}(dt-a\sin^2\theta d\phi)^2+\frac{\sin^2\theta}{\Sigma}\left[(r^2+a^2)d\phi-adt\right]^2\\
    +&\frac{\Sigma}{\Delta}dr^2+\Sigma d\theta^2\,,
\end{split}
\end{equation}
where $\Sigma \coloneqq r^2+a^2\cos^2\theta$ and $\Delta \coloneqq r^2-2Mr+a^2=(r-r_+)(r-r_-)$.
The radii of the outer horizon, $r_+$, and inner horizon, $r_-$, are $r_\pm=M\pm\sqrt{M^2-a^2}$.

The Kerr perturbations are governed by the Teukolsky master equation \cite{Teukolsky:1972my}. 
For a field of spin weight $s$ (where $s=0, \pm1, \pm2$ for scalar, electromagnetic, and gravitational perturbations, respectively), the perturbation $\psi$ can be decomposed as,
\begin{equation}
    \psi(t,r,\theta,\phi) =  e^{-i\omega t}e^{im\phi} {}_{s}S_{\ell m}(\theta; a\omega) R_{\ell m \omega}(r),
\end{equation}
where $\ell$ and $m$ denote the angular and azimuthal mode indices, respectively.
The spin-weighted spheroidal harmonics ${}_{s}S_{\ell m}(\theta; a\omega)$ satisfy the angular Teukolsky equation,
\begin{align}\label{eqn:TklskyAngular}
\begin{split}
    &\frac{1}{\sin\theta} \frac{d}{d\theta}\left(\sin\theta \frac{dS}{d\theta}\right) + \biggl[a^2 \omega^2 \cos^2\theta   - 2a\omega s\cos\theta\\
    &\left.- \frac{(m+s\cos\theta)^2}{\sin^2\theta} +s+ {}_{s}A_{\ell m} \right] S = 0,
\end{split}
\end{align}
with the separation constant ${}_{s}A_{\ell m}$. For $a\omega=0$, the eigenfunctions ${}_{s}S_{\ell m}(\theta, a\omega)$ reduce to the spin-weighted spherical harmonics, and the eigenvalue of the separation constant reduces to ${}_{s}A_{\ell m}=(\ell-s)(\ell+s+1)$. 
The radial function $R_{\ell m\omega}(r)$ satisfies the radial Teukolsky equation,
\begin{align}\label{eqn:TklskyRadial}
\begin{split}
    &\Delta^{-s} \frac{d}{dr}\!\left(\Delta^{s+1} \frac{dR}{dr}\right)\\
    &+\left[\frac{K^2 - 2is(r-M)K}{\Delta}+ 4is\omega r - {}_{s}\lambda_{\ell m}\right] R = {\cal S}_{\rm T},
\end{split}
\end{align}
where ${\cal S}_{\rm T}$ is the source term in the Teukolsky formalism and
\begin{align}
    K&:=(r^2+a^2)\omega -am,\\
    {}_s\lambda_{\ell m}&:= {}_sA_{\ell m}+a^2\omega^2-2am\omega.
\end{align}

In the following, we focus on gravitational perturbations with $s=-2$. 
In the Newman-Penrose formalism, outgoing gravitational radiation is described by the Weyl scalar $\Psi_4$ through,
\begin{equation}
    \Psi_4=-\ddot{h}_+ + i\ddot{h}_\times,
\end{equation}
where the overdot denotes differentiation with respect to $t$, and $h_+$ and $h_\times$ are the ``plus'' and ``cross'' polarisation modes of gravitational waves.

The radial equation \eqref{eqn:TklskyRadial} has a long-ranged potential, which makes direct numerical integration challenging. This is remedied by working instead with the Sasaki-Nakamura (SN) equation \cite{Sasaki:1981kj,Sasaki:1981sx}, which is related to the Teukolsky equation by the Darboux-Chandrasekhar transformation (details in Appendix \ref{app:SN}),
\begin{equation}\label{eqn:SN}
    \left\{\frac{d^2}{dr_*^2} + \left[ \frac{{\cal F}'(r)}{2} - \frac{{\cal F}^2(r)}{4} - {\cal U}(r) \right] \right\}X_{\ell m\omega}(r)={\cal S}_{\rm SN},
\end{equation}
where ${\cal S}_{\rm SN}$ is the source term in the SN formalism and $r_*$ is the tortoise coordinate defined by
\begin{align}
    \frac{dr_*}{dr} &=\frac{r^2+a^2}{\Delta} \,, \label{eqn:drstar}\\
\begin{split}\label{eqn:rstar_explicit}
    r_* &= r
+\frac{2Mr_+}{r_+-r_-}
 \ln\left(\frac{r-r_+}{r_+-r_-}\right)\\
&~~~~~-\frac{2Mr_-}{r_+-r_-}
 \ln\left(\frac{r-r_-}{r_+-r_-}\right)\,.
 \end{split}
\end{align}
In the Schwarzschild limit, $a\rightarrow 0$, the SN equation \eqref{eqn:SN} reduces to the Regge-Wheeler equation \cite{Regge:1957td}.

\subsection{Asymptotics}

In the SN formalism, we define two independent solutions, $X^{\text{in}}, X^{\text{up}}$, to the homogeneous equation, which have the asymptotic forms
\begin{align}
&X^{\text{in}}_{\ell m\omega}\sim
\begin{cases}
    e^{-ik_H r_*}, &\quad r_*\rightarrow -\infty,\\
    A_{\text{in}, \ell m} e^{-i\omega r_*} + 
    A_{\text{out}, \ell m} e^{i\omega r_*}, &\quad r_*\rightarrow +\infty,
\end{cases}\\
&X^{\text{up}}_{\ell m\omega}\sim
\begin{cases}
    B_{\text{in}, \ell m} e^{-i k_H r_*} + 
    B_{\text{out}, \ell m} e^{ik_H r_*}, &\quad r_*\rightarrow -\infty,\\
    e^{i\omega r_*}, &\quad r_*\rightarrow +\infty,
\end{cases}
\end{align}
where $k_H := \omega - m\Omega_H$ is the frequency measured by a co-rotating observer at the horizon and $\Omega_H \coloneqq a/(2Mr_+)$ is the horizon angular velocity.

For a source term that has its compact form  at $r^* = x'$ with $M \ll x' \ll r_{*}$, the Green's function reduces to the following form \cite{Andersson:1996cm}:
\begin{equation}\label{eqn:G_reduced}
    G (r_*, x', \omega) \simeq -\frac{1}{2i \omega} \left[ e^{i \omega (r_* -x')} + \frac{A_{\rm out}}{A_{\rm in}} e^{i \omega (r_* + x')} \right]\,.
\end{equation}
The first term describes free propagation from the source directly to the observer and contributes only to the prompt response. Information about the black hole is completely captured by the reflection coefficient
\begin{equation}
    \mathcal{R}(\omega) \coloneqq \frac{A_{\rm out}(\omega)}{A_{\rm in}(\omega)},
\end{equation}
whose analytic structure in the complex $\omega$ plane determines both the QNM ringing and the tail~\cite{Leaver:1986gd,Andersson:1996cm}.

\subsection{Prompt, QNM and branch-cut contributions}\label{sec:greens}

The retarded time-domain response at the observer is obtained by inverting the Fourier transform,
\begin{equation}\label{eqn:hfull}
    h(u) = \frac{1}{2 \pi} \int_{-\infty + i\varepsilon}^{+\infty + i\varepsilon}  d\omega\, \mathcal{R}(\omega)\, \tilde{T}(\omega)\, e^{-i\omega u},
\end{equation}
where $u:=t-r_*$ is the retarded time, the contour runs above the singularities of the integrand, and $\tilde{T}$ encodes the initial data (Sec.~\ref{sec:initial_data}). For $u>0$ the contour may be closed in the lower half plane, and the integral splits into three physically distinct pieces \cite{Leaver:1986gd,Nollert:1992ifk,Andersson:1996cm},
\begin{equation}\label{eqn:split}
    h(u) = h_{\rm P}(u) + h_{\rm QNM}(u) + h_{\rm tail}(u).
\end{equation}
In this work, we factor out the spin-weighted spheroidal harmonics ${}_{s}S_{\ell m} (a \omega, \theta)$ to isolate the observer-independent, i.e., $\theta$-independent, features of the branch-cut contribution. 
We leave a detailed investigation of these observer-dependent effects for future work.

\emph{(i) Prompt response.} The contribution of the large semicircle at $|\omega| \to \infty$ reflects the high-frequency propagation of the initial data. It is non-zero only over a finite interval of retarded time set by the light travel time between the support of the initial data and the observer, and it is insensitive to the black hole parameters. Together with the first term in equation \eqref{eqn:G_reduced}, it forms the direct, or ``prompt'' part of the signal.
In this paper, we only focus on a signal after the onset of QNM and branch-cut contribution, and we do not consider the prompt signal.

\emph{(ii) Quasinormal modes.} The amplitude $A_{\rm in}(\omega)$ possesses an infinite set of zeros in the lower half plane, located at the QNM frequencies $\omega_{\ell m n}$ with $\Im(\omega_{\ell m n})<0$. These are simple poles of $\mathcal{R}(\omega)$, and their residues define the excitation factors $E_{\ell m n}$ \cite{Leaver:1986gd,Berti_2006,Zhang:2013ksa}. Summing over the residues gives
\begin{equation}\label{eqn:h_qnm}
    h_{\rm QNM}(u) = \sum_{n} E_{\ell m n}\, \tilde{T}(\omega_{\ell m n})\, e^{-i \omega_{\ell m n} u},
\end{equation}
such that the mode amplitude factorises into a source-independent excitation factor, and the frequency-domain source term evaluated at the mode frequencies. 
The Kerr QNM spectrum has two branches, the prograde modes with $\Re(\omega)>0$ and the retrograde (mirror) modes.
In our analysis we retain both, since the retrograde branch is not negligible at early times (see e.g., Ref.~\cite{Dhani:2020nik}).

\emph{(iii) Branch cut.} In addition to the poles, $\mathcal{R}(\omega)$ has a branch point at $\omega=0$. The cut is typically taken along the negative imaginary axis, $\omega = -iy$ with $y>0$, then the remaining contribution is
\begin{equation}\label{eqn:h_tail}
    h_{\rm tail}(u) = -\frac{1}{2\pi} \int_{0}^{\infty} \!\! dy\, \Delta\mathcal{R}(-iy)\, \tilde{T}(-iy)\, e^{-y u},
\end{equation}
where $\Delta\mathcal{R} \coloneqq \lim_{\delta_{\rm R} \to 0^+} \mathcal{R}(\delta_{\rm R} -iy) - \mathcal{R}(-\delta_{\rm R} -iy)$ is the discontinuity across the cut. 
Expanding $\Delta\mathcal{R}$ about $y=0$ and performing the integral leads to $h_{\rm tail} \sim u^{-2}$ at late times \cite{Rosato:2025rtr}. 
Here, we do not focus on this asymptotic form, since we are interested in the early and intermediate time tail. 
Equation \eqref{eqn:h_tail} defines the branch cut contribution at all times, and we use this to look at earlier time behaviour.

It is convenient to separate the black hole response from the initial data. We define the response function
\begin{align}
    h_G(u) :=& \frac{1}{2\pi}\int d\omega e^{-i\omega u}\mathcal{R}(\omega),\label{eqn:h_G}\\
    h_T(u):=&\frac{1}{2\pi}\int d\omega e^{-i\omega u} \tilde{T}(\omega).
\end{align}
Then equation \eqref{eqn:hfull} is equivalent to the convolution,
\begin{equation}\label{eqn:conv}
    h(u) = \left( h_G * h_T \right)(u) = \int du'\, h_G(u-u')\, h_T(u').
\end{equation}
Especially, at late times, $u > u_*$, the waveform $h_{{\rm G}} (u)$ can be decomposed into the two parts:
\begin{equation}
h_{{\rm G}} (u) = h_{{\rm G}, \rm QNM} (u) + h_{{\rm G}, \rm tail} (u)\,,
\end{equation}
where $u_*$ is a value associated with the excitation region of Kerr QNMs \cite{Oshita:2026vxh}.
Since the decomposition \eqref{eqn:split} of $h$ is linear, each component of $h_G$ may be convolved with the same source independently, to give the excitation of each component separately. 
\begin{equation}\label{eqn:convsplit}
    h_{\rm QNM} = h_{{\rm G},{\rm QNM}} * h_T, \qquad h_{\rm tail} = h_{{\rm G},{\rm tail}} * h_T.
\end{equation}

\subsection{Computation of \texorpdfstring{$h_{{\rm G}, \rm QNM}$}{h\_G,QNM} and \texorpdfstring{$h_{{\rm G}, \rm tail}$}{h\_G,tail}}

The QNM part from the reflection sector, $h_{{\rm G}, \rm QNM} (u)$, is
\begin{equation}
    h_{{\rm G}, \rm QNM} (u) = -i \sum_{\sigma, n} E_{\ell m n}^{(\sigma)} \exp \left( - i \omega_{\ell mn}^{(\sigma)} u \right)\,,
\end{equation}
where $\sigma = +$ and $-$ stands for the prograde and retrograde modes, respectively.
$E_{\ell m n}^{(\sigma)}$ is the QNM excitation factor defined as
\begin{equation}
    E_{\ell m n}^{(\sigma)} \coloneqq \left. \frac{A_{\rm out}(\omega)}{A_{\rm in}' (\omega)} \right|_{\omega = \omega_{\ell m n}^{(\sigma)}}\,,
\end{equation}
where the prime denotes the derivative with respect to $\omega$.
The values of $A_{\rm out}$ and $A_{\rm in}$ are computed by the MST method \cite{Mano:1996vt,Zhang:2013ksa}.

The branch-cut contribution, $h_{{\rm G},\rm tail} (u)$, is also computed by the MST method.
The branch-cut waveform can be obtained via integration along the branch cut located on the negative-imaginary axis:
\begin{align}
\begin{split}
    h_{{\rm G}, \rm tail} (u) &= \lim_{\delta_{\rm R} \to 0} i \int_{- \infty}^0  d \omega_{\rm I} \left[ {\cal R} (\delta_{\rm R} + i \omega_{\rm I}) e^{-i (\delta_{\rm R} + i \omega_{\rm I}) u} \right.\\
    &\left. ~~~~~~+ {\cal R} (-\delta_{\rm R} + i \omega_{\rm I}) e^{-i (- \delta_{\rm R} + i \omega_{\rm I}) u} \right] \,.
    \label{tail_MST}
\end{split}
\end{align}
In the numerical evaluation of Eq.~\eqref{tail_MST}, the regulator $\delta_{\rm R} (>0)$ must be kept finite and nonzero. 
In addition, the integration along the negative imaginary-frequency axis must be truncated at a finite lower bound, $\omega_{\rm I}=\omega_{\rm I}^{\rm min}$, chosen sufficiently deep in the region where the integration result is sufficiently convergent. 
The choice of $\delta_{\rm R}$ and $\omega_{\rm I}^{\rm min}$ is discussed in Appendix~\ref{app:tail}.

\section{Response to initial data}\label{sec:response_initial}
\subsection{Initial data}\label{sec:initial_data}
Following \cite{Andersson:1996cm, Berti_2006}, we take the initial data to be specified by a function $T(x)$ whose transform is given by 
\begin{equation}
    \tilde{T}(\omega) = \int_{-\infty}^{+\infty} dx'\, \tilde{S}(x', \omega)\, e^{i\omega x'},
\end{equation}
where $\tilde{S}(r_*,\omega) := -{\cal S}_{\rm SN}(r_*, \omega) / (2 \pi i \omega)$. 
To specify the initial data explicitly, let
$X_{\ell m}(t,r_*)$ denote the time-domain Sasaki-Nakamura variable.
Following Ref.~\cite{Berti_2006}, we consider stationary initial data, localised in the far zone at $t=0$:
\begin{align}\label{eqn:cauchy_data}
    &X_{\ell m}(0,r_*)=-\frac{2\pi}{\cal N} e^{-(x-x_0)^2/\sigma^2}\, e^{-i\omega_0 (x-x_0)},\\
    &\left.\partial_t X_{\ell m}(t,r_*)\right|_{t=0}=0.
\end{align}
We consider a fixed $\sigma$ and a large $x_0$, for which the source term is approximated as ${\cal S}_{\rm SN} \simeq i \omega X_{\ell m} (0,r_*)$.
For the initial data, we have
\begin{equation}\label{eqn:gaussian}
    \tilde{S}(x) = \frac{1}{\mathcal{N}}\, e^{-(x-x_0)^2/\sigma^2}\, e^{-i\omega_0 (x-x_0)},
\end{equation}
normalised such that the envelope carries unit area,
\begin{equation}\label{eqn:norm}
    \mathcal{N} = \int_{-\infty}^{\infty} dx\, \left| e^{-(x-x_0)^2/\sigma^2} \right| = \sigma\sqrt{\pi}.
\end{equation}
Finally, we have
\begin{equation}\label{eqn:Ttilde}
    \tilde{T}(\omega) = e^{i\omega x_0}\, e^{-\sigma^2(\omega-\omega_0)^2/4},
\end{equation}
which is a Gaussian of width $2/\sigma$ in frequency, centred on $\omega = \omega_0$. 

Throughout this paper, we fix $\sigma=1$, and vary the central frequency over $2M\omega_0 \in \{0, 0.5, \dots, 5.0\}$. We consider spins $\chi \in \{0, 0.1, \dots, 0.9, 0.99, 0.999, 0.99999\}$. 
The initial location of the wave packet $r_* = x_0$ is absorbed into the retarded time in our setup, but as mentioned, our description of the source term is verified only when the wave packet is initially localised at far zone.
The convolution \eqref{eqn:conv} is evaluated directly in the time domain on a uniform grid of spacing $\delta u/(2M) = 10^{-2}$ spanning $u/(2M) \in [-5, 200]\,$. For the QNM sum in \eqref{eqn:h_qnm} we retain more than $100$ overtones in total (see Table~II in Ref.~\cite{Oshita:2026vxh}). 

\subsection{Start time of the ringdown}
As discussed in Sec.~\ref{sec:greens}, the QNM and branch-cut contributions together account for the entire signal once the prompt response has decayed at $u > u_0$. 
We compare $h_{{\rm G}} * h_T$ with the full waveform $h(u)$ to identify the starting time $u_0$. 
We evaluate the start time, for a given tolerance $\epsilon$, as
\begin{equation}\label{eqn:starttime}
    u_0(\epsilon) := \min\left\{ u > u_{\rm min} : \left| \frac{\left|h_{\rm QNM}(u)+h_{\rm tail}(u)\right|}{\left|h(u)\right|} - 1 \right| < \epsilon \right\},
\end{equation}
with $u_{\rm min} = -5M$, and all three quantities, $h_{\rm QNM}$, $h_{\rm tail}$ and $h$, computed by convolving with the same initial data. The start time $u_0$ is the earliest time at which a linear description of the ringdown, including the non-oscillatory tail, becomes accurate (within the $\epsilon$ tolerance). 
We emphasise that this procedure is the ringdown reconstruction (see e.g., \cite{Oshita:2024wgt}) rather than a fit. 
Once the black hole parameters and initial data
are fixed, the QNM frequencies and residues, the branch-cut
contribution, and their amplitudes and phases are all determined independently from the Green's function. 
Hence, no free parameters are adjusted to reproduce $h$ from the convolution. 
Consequently, agreement at $u\geq u_0$ cannot be attributed to overfitting, and increasing the number of QNMs in our analysis constitutes a convergence test of the ringdown reconstruction rather than an increase in fitting freedom.

We give our results for $\epsilon=0.01$ in the main text, and demonstrate in Appendix \ref{app:tol} that the qualitative behaviour is unchanged for $\epsilon=0.05,$ and $\epsilon=0.001$. Note that $u_0$ also depends on the spin, $a$, and the parameters of the initial data, e.g., $\omega_0$, so that the amplitudes reported below are compared at a different retarded time for each configuration. 
This is intentional, since the physically meaningful comparison is between the two components at the moment the ringdown analysis would begin.

The central quantities of this paper are then
\begin{equation}
    \mathcal{Q}(\chi,\omega_0):=\frac{\left| h_{\rm QNM}(u_0) \right|}{\left| h_{\rm tail}(u_0) \right|},
\end{equation}
as well as the two amplitudes separately, which we compare to understand the relative enhancement or suppression of each component of the signal.
\section{Results}\label{sec:results}
\subsection{The response}
\begin{figure}
    \centering
    \includegraphics[width=\columnwidth]{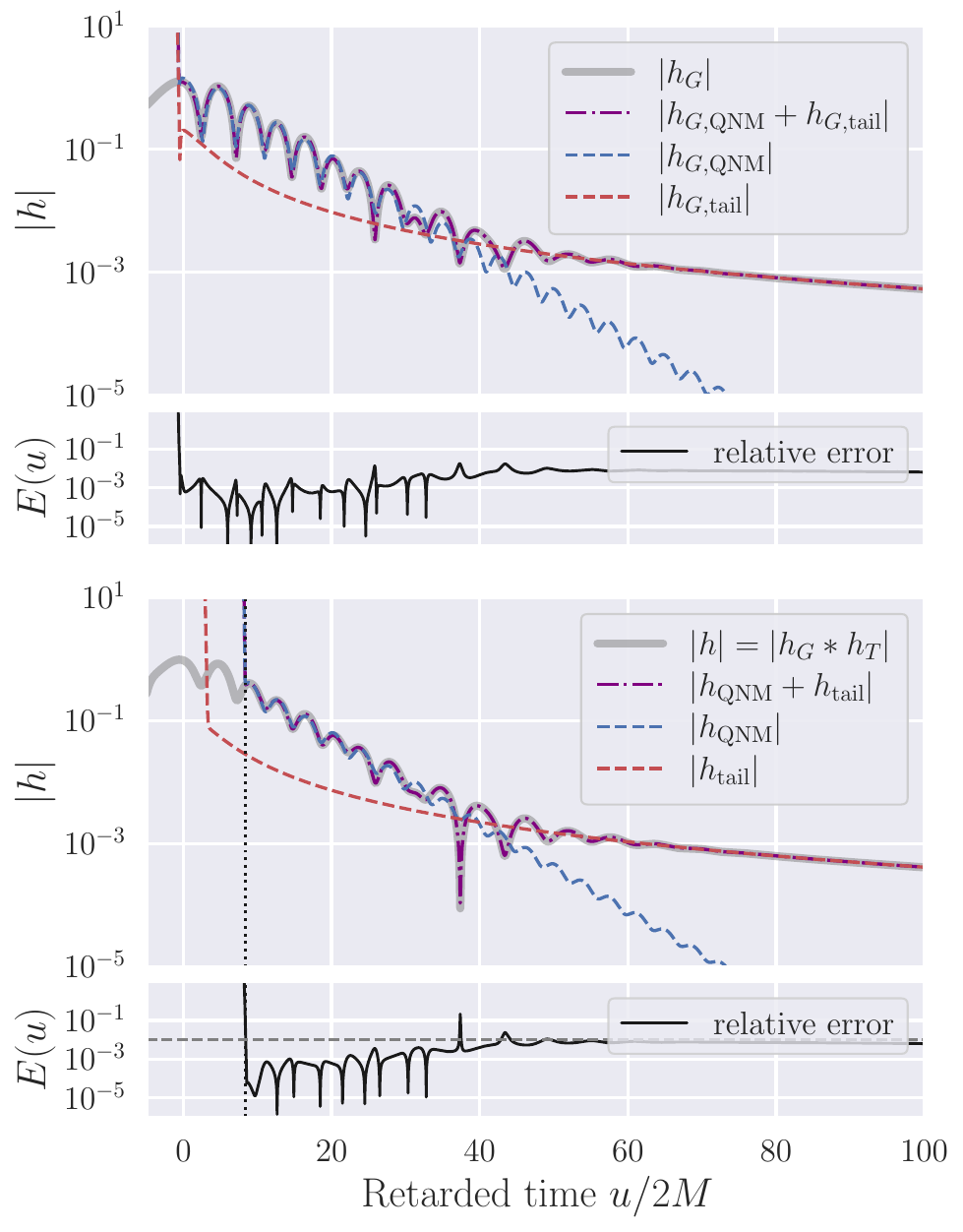}
    \caption{Decomposition of the $\ell=m=2$ response of a Kerr black hole with $\chi=0.7$. \emph{Top:} the impulse response $h_G$ of equation \eqref{eqn:h_G}, as well as the QNM and branch-cut parts and their sum. \emph{Bottom:} the same quantities as above convolved with Gaussian initial data ($\sigma=1, x_0=0, \omega_0=1$), i.e. the physical response $h$. In both panels the thick grey curve is the full response. Below each panel we show the relative error, $E(u)=\bigl||h_{\mathrm{QNM}}+h_{\mathrm{tail}}|/|h|-1\bigr|$, which measures how well the QNM plus tail description reproduces the full response. The vertical dotted line marks the start time $u_0$ (as in equation \eqref{eqn:starttime} with $\epsilon=0.01$) at which the prompt response becomes negligible and the QNM and tail sum approximate the full response well, while the horizontal dashed line marks the threshold $\epsilon$.} 
    \label{fig:waveforms}
\end{figure}

Figure \ref{fig:waveforms} shows the full response of a $\chi=0.7$ black hole to a Gaussian wave packet with $\sigma=1,\omega_0=1$, together with its QNM and tail components, before and after convolution with the initial data.
Below each panel, we also display the relative error $E(u)$, defined as
\begin{equation}
    E(u):=\left|\frac{|h_{\mathrm{QNM}}+h_{\mathrm{tail}}|}{|h|}-1\right|.
\end{equation}
The qualitative structure is generic across the parameter space. 
At earlier times, the sum of QNMs and branch-cut contribution exhibits the divergent behaviour, where the QNM-and-tail decomposition fails due to the causality of wave scattering.
At $u = u_0$, the tail and QNM sum approximate the full signal well. Beyond the start time, the QNM signal dominates but decays exponentially, while the tail decays as a power law, eventually taking over around $u\approx40$. At late times the QNM has decayed away, leaving only power-law tail.

\subsection{Spin dependence of early-time tail}
We begin by comparing the tail waveforms across spin, before turning to their amplitudes relative to the QNM contributions. 

The branch-cut integral of equation~\eqref{tail_MST} is evaluated along the cut down to $\omega_\mathrm{I}^{\mathrm{min}}$, and grows exponentially for $u<0$. The physical tail emerges once this contribution has decayed, and we find the onset to be spin-dependent. 
We shift the retarded time to account for the spin-dependent offset in the tortoise coordinate and compare the waveforms as functions of
\begin{equation}
    \tilde{u} \equiv u-2M\ln\!\left(1-\chi^{2}\right).
    \label{eqn:ushift}
\end{equation}
Indeed, expanding the tortoise coordinate defined in Eq.~\eqref{eqn:rstar_explicit} at spatial infinity yields
\begin{equation}
    r_* = r+2M\ln\left(\frac{r}{2M}\right)-M\ln(1-\chi^2)+O(r^{-1}),
\end{equation}
revealing a spin-dependent constant offset, $\delta \coloneqq -M\ln(1-\chi^2)$, that induces a time shift of twice this amount in the inverse Fourier transform of the reflection amplitude $A^{\rm out}/A^{\rm in}$ \cite{Oshita:2024wgt}. 
Under $r_*\to r_*+\delta$, the asymptotic amplitudes of the radial solution rotate by opposite phases, $A_{\mathrm{in}} \to A_{\mathrm{in}}e^{+i\omega\delta}$ and $A_{\mathrm{out}} \to A_{\mathrm{out}}e^{-i\omega\delta}$. Hence their ratio, which is the quantity that is Fourier transformed to obtain the tail, acquires a factor of $e^{-2i\omega\delta}$, and the inverse transform is displaced by $2\delta$.

Accordingly, when considering waveforms obtained without convolution with the source, we compare them as functions of the shifted retarded time defined in Eq.~\eqref{eqn:ushift}.
We emphasise that when convolved with the source to reconstruct the full waveform, the corresponding phase shifts cancel, leaving the physical waveform independent of the choice of tortoise-coordinate convention.

\begin{figure}
    \centering
    \includegraphics[width=\columnwidth]{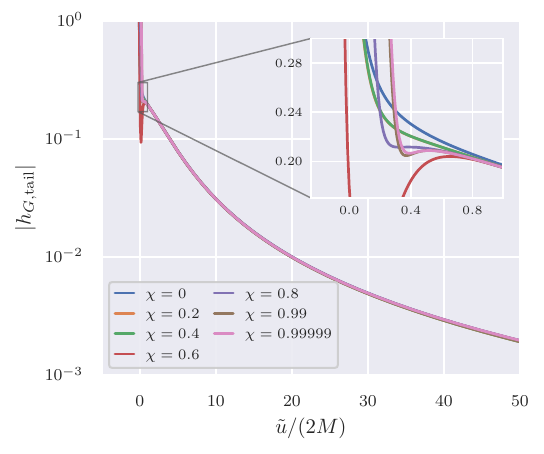}
    \caption{The branch-cut contribution $|h_{{\rm G},\mathrm{tail}}|$ to the $\ell=m=2$ ringdown for a range of spin parameters $\chi$, plotted against the shifted retarded time $\tilde{u}=u-2M\ln(1-\chi^{2})$. The onset of the tail occurs at a spin-dependent retarded time; once shifted to remove this displacement, the waveforms are relatively insensitive to the spin parameter, agreeing with each other to better than $2\%$ across the range $1\leq \tilde{u}/(2M) \leq 50$, such that the curves plotted are almost indistinguishable. The inset shows the onset region, where the spin-dependence is most prominent. The deviation between spins in the early-time tail is confined to $0\lesssim\tilde{u}/(2M)\lesssim 1$. For sake of clarity, we show only half of the spin parameters considered in the rest of the paper, but the qualitative results are the same when we include all the spins.}
    \label{fig:early_tail}
\end{figure}

Figure~\ref{fig:early_tail} shows the relatively weak spin-dependence of the early-time tail, up to a time shift. Plotted as a function of the shifted retarded time $\tilde{u}$, the waveforms collapse onto a single curve. The main deviations between the curves are confined to a narrow window around the onset of the tail, as shown in the inset. This is not an automatic consequence of the known late-time behaviour. The asymptotic power-law decay of the tail is known to exhibit a degree of universality \cite{Rosato:2025rtr}, but it is non-trivial how the universality can be broken at early times, where the branch-cut integral is not dominated by the neighbourhood of $\omega=0$, and the detailed structure of $A_{\mathrm{out}}/A_{\mathrm{in}}$ along the cut may in principle depend on $\chi$. 

Despite the insensitivity of the early-time tail to the black hole spin, we show in the following that the ratio $\mathcal{Q}$ is still spin-dependent. This spin-dependence follows primarily from the QNM contribution, as well as the definition of the start time $u_0$ we have adopted.

\begin{figure}
    \centering
    \includegraphics[width=\columnwidth]{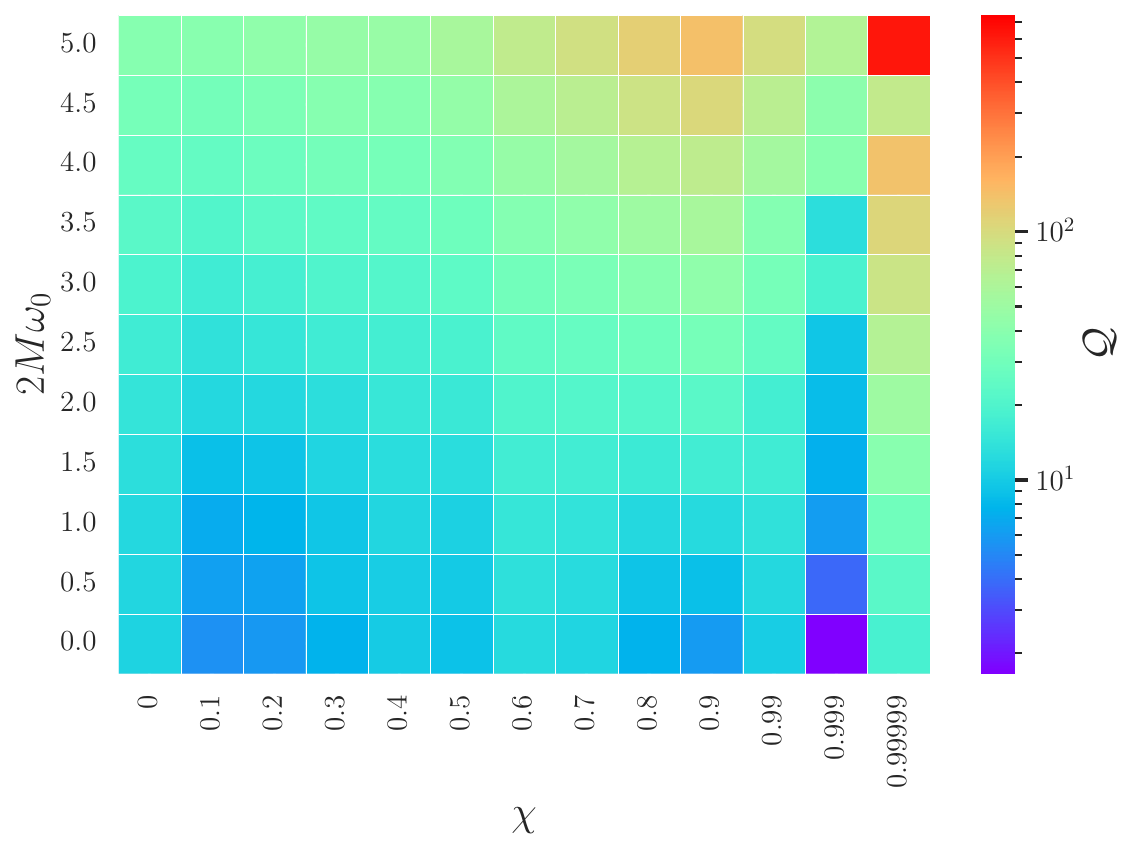}
    \caption{The ratio $\mathcal{Q}(\chi,\omega_0)$ of QNM to tail amplitude, evaluated at the start time $u_0$, with tolerance $\epsilon=0.01$, and Gaussian initial data with $\sigma=1, x_0=0$.}
    \label{fig:ratios}
\end{figure}

In Figure \ref{fig:ratios} we show $\mathcal{Q}(\chi,\omega_0)$ for $\epsilon=0.01$ across the full grid of spins and central frequencies. We find that $\mathcal{Q}>1$ everywhere: at the moment the ringdown becomes describable by the QNM sum and tail contributions, the signal is dominated by QNMs, typically by one to two orders of magnitude in amplitude. For most of the parameter space, $\mathcal{Q}$ lies in the range $5 \lesssim \mathcal{Q} \lesssim 100$, rising to a maximum of around $\mathcal{Q} \gtrsim 10^2$ in the upper-right corner, at large spin and large $\omega_0$.

At fixed $\omega_0$, $\mathcal{Q}$ has two qualitatively distinct behaviours depending on the value of $\omega_0$. For $2M\omega_0\gtrsim 2$, $\mathcal{Q}$ increases monotonically with spin. However for $2M\omega_0\lesssim 1.5$, the ratio increases before peaking near $\chi \simeq 0.6-0.7$, and then decreases again. 

\begin{figure*}
    \centering
    \includegraphics[width=\textwidth]{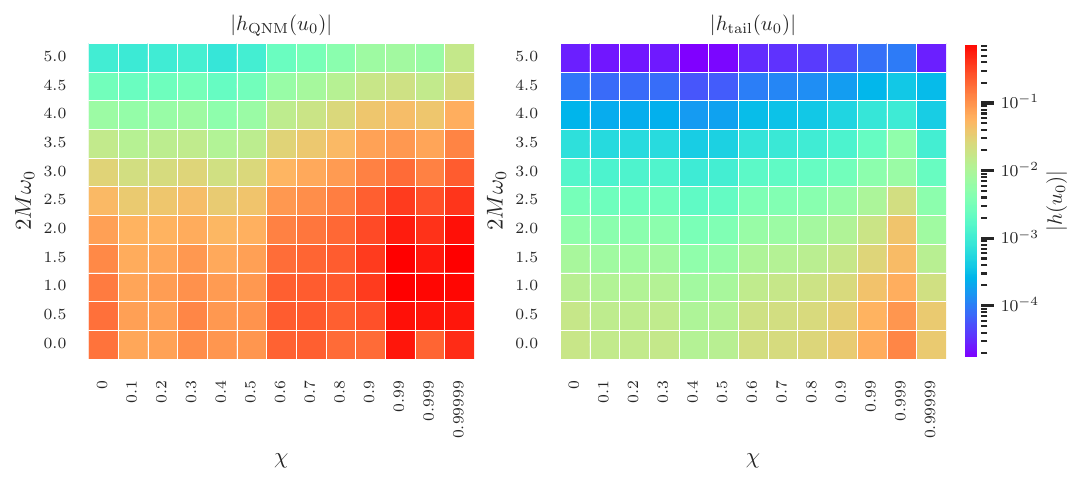}
    \caption{\emph{Left:} the QNM amplitude at the start time. \emph{Right:} the tail amplitude at the start time. For both plots the start time is found using a tolerance of $\epsilon=0.01$. }
    \label{fig:amplitudes}
\end{figure*}

Figure \ref{fig:amplitudes} separately shows the QNM sum and tail amplitudes across the parameter space $(\chi,\omega_0)$. Both the QNM and tail amplitude at $u_0$ grow with spin, but the QNM amplitude is considerably higher as $5 \lesssim {\cal Q} \lesssim 100$.
The natural expectation is that as the black hole spins up, the corotating $\ell=m=2$ modes become long lived, $|\Im(\omega_{22n})|\to0,$ and are most easily excited, such that a larger fraction of the perturbation energy is stored in the quasinormal modes, while the branch cut depends on the asymptotic behaviour, and is less sensitive to the near-horizon geometry.
We find that for higher spins but smaller than the Thorne limit, $0.7 \lesssim \chi \lesssim 0.998$, the branch-cut amplitude is amplified by the BH spin.
On the other hand, for a fixed spin, the QNM amplitude is slightly more stable as the frequency of the initial data is increased. However the tail amplitude drops off more quickly as $\omega_0$ is increased; this is in agreement with previous studies which have explained the origin of tails as being generated by low frequency waves \cite{Cardoso:2024jme, Islam:2024vro}.

\begin{figure}
    \centering
    \includegraphics[width=\columnwidth]{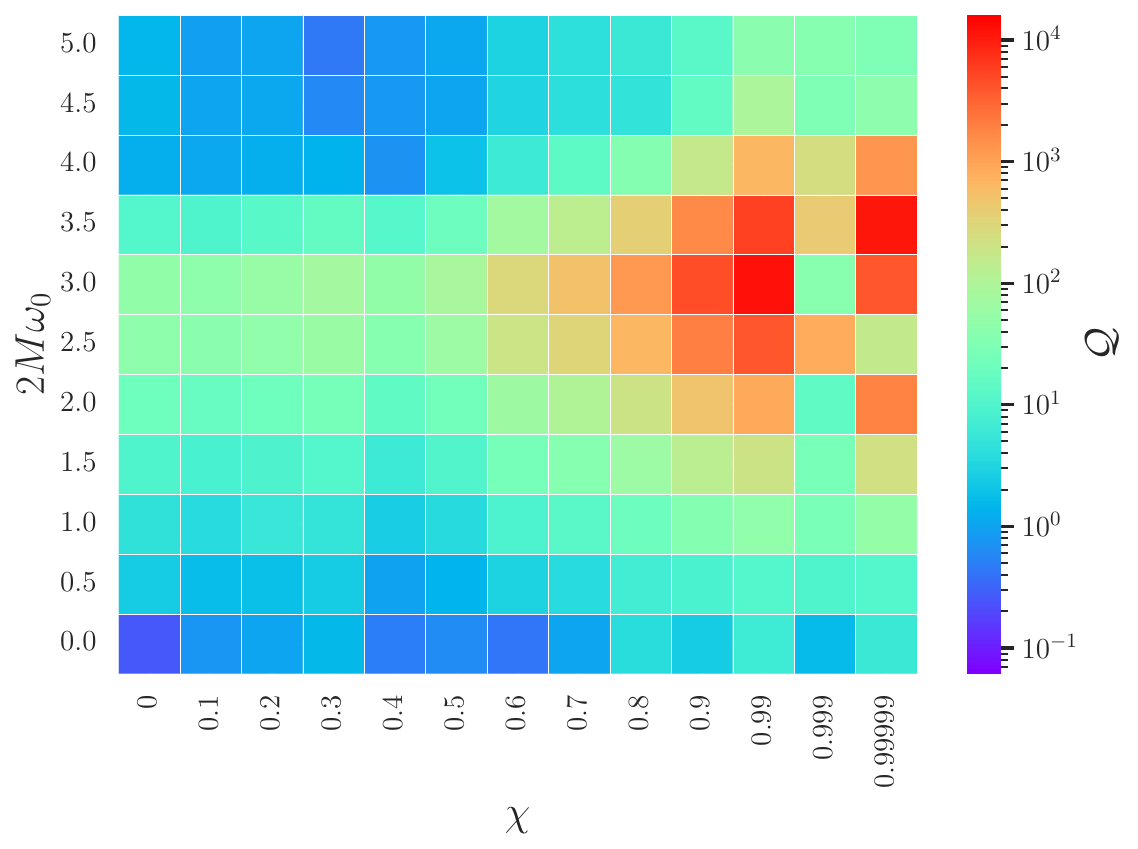}
    \caption{The ratio $\mathcal{Q}(\chi,\omega_0)$ of QNM to tail amplitude, evaluated at the start time $u_0$, with tolerance $\epsilon=0.01$, and Gaussian initial data with $\sigma=2, x_0=0$. In contrast with Figure \ref{fig:ratios}, here there is resonant excitation of QNMs, leading to an enhanced ratio across all spins considered.}
    \label{fig:ratios_sig_2}
\end{figure}

In Figure \ref{fig:ratios_sig_2}, we show the resonant excitation of QNMs for Gaussian initial data with $\sigma=2$. For this value of $\sigma$, QNM amplitudes are enhanced around $2M\omega_0\simeq3$, which leads to an enhanced ratio $\mathcal{Q}$. The QNM enhancement is strongly spin-dependent: at $\chi=0$ the ratio remains small, $\mathcal{Q}\lesssim 10$ across the whole range of $\omega_0$, whereas for $\chi\gtrsim 0.9$, the peak ratio is larger by up to three orders of magnitude. Therefore, the tail is least significant relative to the QNM contribution, in the regime of rapid rotation and resonant driving. In contrast, the $2M\omega_0=0$ row shows $\mathcal{Q}<1$ at several spins, indicating that for a non-oscillatory initial pulse the tail amplitude at $u_0$ exceeds the QNM amplitude. Indeed, this demonstrates that the early-time tail cannot always be assumed subdominant, whether it is or not is strongly dependent on both the initial data and the black hole parameters.

\section{Discussion}\label{sec:disc}
\subsection{Implications for ringdown modelling}
For the class of initial data considered here, characterised by a typical frequency $\omega_0$, the tail contribution is between one and two orders of magnitude below the QNM content at the start of ringdown for all spins, even with $2M\omega_0 \sim 1$. 
The fractional error incurred by omitting the tail of $\sim1-10\%$ is not insignificant. 
For the waveforms considered here, the value of $\mathcal{Q}$ at $u_0$ appears to provide an approximate upper bound on the QNM-to-tail amplitude ratio at the subsequent times. 
The tail fraction is therefore generally expected to become larger at later times, as QNMs exponentially decay whereas the branch-cut contribution decay in power law.\footnote{We do not, however, expect strictly monotonic behaviour at early times, since interference among the simultaneously excited QNMs, including partial destructive interference, can produce non-trivial variations in the instantaneous amplitude ratio.}
If the branch-cut contribution is omitted from a ringdown model, it may be partially absorbed into the fitted QNM amplitudes and phases.
Therefore, our results suggest that it is not justifiable to omit tails in ringdown analysis on the basis that they are negligible. 
Whether this leads to appreciable parameter biases requires a dedicated inference analysis.
It is interesting to assess how crucial it is to include tails in ringdown modelling for the next-generation GW detectors.
Whether our findings and conclusions based on the wave-scattering experiment are applicable to more realistic systems, such as extreme-mass-ratio coalescences, remains to be established through analyses that go beyond our setup in this work. 
We leave such an extension to future work.

\subsection{Astrophysical initial data and eccentric encounters}
We have found that the early-time tail component is strongest in the limit of small frequency perturbations. 
This provides insight into which astrophysical configurations may depart from the QNM-dominated picture found here.

The Gaussian initial data we considered are spectrally narrow and centred away from the origin, which is unfavourable for tail excitation. A quasi-circular inspiral deposits its energy in a narrow band around twice the orbital frequency and should behave like our large-$\omega_0$ cases. However, highly eccentric binaries, dynamical captures, and hyperbolic encounters have the opposite behaviour. An eccentric orbit radiates in sharp bursts at periapsis, but spends most of its time at large separation near the apoapsis, where the emission is at much lower frequencies \cite{Islam:2024vro}. 
In this case, the system deposits more power at $\omega \sim 0$ than a quasi-circular inspiral of comparable energy. This is the mechanism identified by De Amicis \textit{et al.} \cite{DeAmicis:2024not} where they showed that the tail amplitude in non-circular configurations is enhanced by orders of magnitude relative to the quasi-circular case, with the dominant contributions coming from when the particle is far from the black hole and moving slowly shortly before plunge. Related enhancements have since been seen in fully nonlinear merger simulations \cite{DeAmicis:2024eoy,Ma:2024hzq} and analysed in eccentric binaries \cite{Islam:2025wci}. More broadly, tails are also generic in dynamical spacetimes \cite{Cardoso:2024jme}.

Our results extend this picture by revealing how black hole spin affects the relative importance of the branch-cut and QNM contributions, particularly around the onset of ringdown.
The configurations most likely to produce a tail-contaminated ringdown are remnants with low-frequency dominated, highly eccentric or hyperbolic encounters. 
Remnants of quasi-circular mergers, which typically have spins around $\chi\sim0.7$ and are predominantly excited at relatively high frequencies, lie in a regime where QNM dominance is expected to be robust.

\vskip 1cm
\acknowledgments
A.G. is supported by King’s College London through an NMES funded studentship and by the Young International Researcher Invitation Program at Yukawa Institute for Theoretical Physics. 
N.O. was supported by Japan Society for the Promotion of Science (JSPS) KAKENHI Grant No.~JP26K17142.

\appendix
\section{Sasaki-Nakamura Formalism}\label{app:SN}
Here we summarise the main parts of the Sasaki-Nakamura (SN) formalism used to compute the radial scattering amplitudes and QNM excitation factors. The aim of the results presented here is to facilitate computation. For a more detailed discussion we refer the reader to the original papers  \cite{Sasaki:1981kj,Sasaki:1981sx}.

The Teukolsky equation provides the natural gauge invariant description of perturbations of the Kerr solution, while the SN transformation maps the radial Teukolsky equation for gravitational perturbations into an equation with a short-ranged potential. This makes the extraction of asymptotic amplitudes more numerically stable. 

Starting from \eqref{eqn:TklskyRadial}, with $s=-2$, there is a two-step procedure to arrive at \eqref{eqn:SN}. First, we transform the radial Teukolsky wavefunction $R(r)$ to the SN wavefunction $\xi$ through
\begin{equation}
    \xi = \frac{\sqrt{r^2+a^2}}{\Delta}\left(\alpha (r) R(r) + \frac{\beta(r)}{\Delta}R'(r)\right),
\end{equation}
where the prime denotes differentiation with respect to $r$. The functions $\alpha$ and $\beta$ are determined by three auxiliary functions $f, g,$ and $h$, which are otherwise arbitrary. Following \cite{Sasaki:1981sx}, we set $f=h=1$ and $g=(r^2+a^2)/r^2$, since this choice ensures that the SN equation reduces to the Regge-Wheeler equation in the Schwarzschild limit $a\rightarrow0$. With this specification, the functions $\alpha$ and $\beta$ take the forms,
\begin{align}
    \alpha &= -\frac{iK}{\Delta^2}\beta + 3iK' + \lambda + \frac{6\Delta}{r^2},\\
    \beta &= \Delta\left(-2iK + \Delta' - \frac{4\Delta}{r}\right).
\end{align}
The definitions of $K,\Delta,$ and $\lambda$ are given in the main text in section \ref{sec:pert_Kerr}. With these choices, the SN wavefunction $\xi$ satisfies
\begin{equation}
    \frac{d^2\xi}{dr_*^2}-\mathcal{F}\frac{d\xi}{dr_*}-\mathcal{U\xi} = S,
\end{equation}
where the functions $\mathcal{F}$ and $\mathcal{U}$ are given by
\begin{align}
    \mathcal{F}&=\frac{\Delta}{r^2+a^2}F, \quad F=\frac{\gamma'}{\gamma},\\
    \gamma &= \alpha\left(\alpha + \frac{\beta'}{\Delta}\right) - \frac{\beta}{\Delta}\left(\alpha' + \frac{\beta}{\Delta^2} V\right),\\
    \mathcal{U}&=\frac{\Delta U}{\left(r^2+a^2\right)^2}+G^2 + \frac{dG}{dr_*} - \frac{\Delta GF}{r^2+a^2}, \\
    U &= V + \frac{\Delta^2}{\beta}\left[\left(2\alpha + \frac{\beta'}{\Delta}\right)'-\frac{\gamma'}{\gamma}\left(\alpha + \frac{\beta'}{\Delta}\right)\right],\\
    G &= -\frac{\Delta'}{r^2+a^2} + \frac{r\Delta}{(r^2+a^2)^2}.
\end{align}
In this form, $\mathcal{F}$ and $\mathcal{U}$ are already short-ranged. However, this equation is made even simpler by a further transformation of the wavefunction,
\begin{equation}
\begin{split}
    \xi &= X\exp\left[\int\frac{\mathcal{F}}{2}dr_*\right]= X\sqrt{\gamma}.
\end{split}
\end{equation}
This leads to the form of the SN equation \eqref{eqn:SN}, as quoted in the main text,
\begin{equation}
\begin{split}
    \frac{d^2X}{dr_*^2} + \left[ \frac{{\cal F}'}{2} - \frac{{\cal F}^2}{4} - {\cal U} \right] X &= \frac{S}{\sqrt{\gamma}}={\cal S}_{\rm SN}.
\end{split}
\end{equation}

\section{Computation of \texorpdfstring{$h_{{\rm G}, \rm tail} (u)$}{h\_G,tail}}
\label{app:tail}
\begin{figure}[t]
    \centering
    \includegraphics[width=\columnwidth]{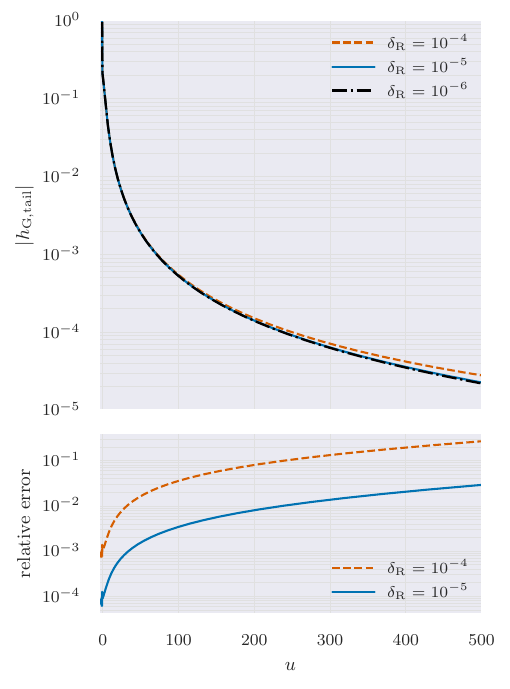}
    \caption{\emph{Top:} The absolute amplitude of $h_{{\rm G},\rm tail}$ ($\chi = 0.8$ and $\ell =m =2$) for $\delta_{\rm R} = 10^{-4}$ (orange dashed), $10^{-5}$ (blue solid) and $10^{-6}$ (black dot-dashed).
    \emph{Bottom:} Relative differences in $|h_{{\rm G},\rm tail}|$ between $\delta_{\rm R} = 10^{-4}$ and $10^{-6}$ (orange dashed) and between $\delta_{\rm R} = 10^{-5}$ and $10^{-6}$ (blue solid).
    } 
    \label{fig:app_convergence_tail}
\end{figure}

Here we compute the branch-cut contribution $h_{{\rm G},\rm tail}$ by using
\begin{align}
\begin{split}
    h_{{\rm G}, \rm tail} (u) &= i \int_{\omega_{\rm min}}^{-\delta_{\rm R}}  d \omega_{\rm I} \left[ {\cal R} (\delta_{\rm R} + i \omega_{\rm I}) e^{-i (\delta_{\rm R} + i \omega_{\rm I}) u} \right.\\
    &\left. ~~~~~~+ {\cal R} (-\delta_{\rm R} + i \omega_{\rm I}) e^{-i (- \delta_{\rm R} + i \omega_{\rm I}) u} \right] \,,
    \label{app_tail_MST}
\end{split}
\end{align}
with $0 < \delta_{R} \ll 1$.
FIG.~\ref{fig:app_convergence_tail} shows the comparison among $h_{{\rm G},\rm tail}$ for $\delta_{\rm R} = 10^{-6}$, $10^{-5}$, and $10^{-4}$.
We confirm the convergence of the tail amplitude $h_{{\rm G},\rm tail}$ for $\delta_{\rm R} \sim 10^{-5}$.
In the main text, we set $\delta_{\rm R} = 10^{-5}$ for $\chi > 0$.
On the other hand, we found that our MST code is unstable near the branch cut for $\chi = 0$.
We therefore compute the branch-cut contribution at $\omega = \delta_{\rm R} (1-i)+ s (0.1  - 15 i)$ with $0 \leq s \leq 1$ for $\chi = 0$.
We set $2M\omega_{\rm min} = -15$ based on Ref.~\cite{Oshita:2026vxh}. 

\section{Dependence on start-time criterion}\label{app:tol}
Here we demonstrate that our findings are robust under small variations of the start time tolerance used in equation \eqref{eqn:starttime}. The tolerance $\epsilon$ fixes how closely the QNM and tail sum must reproduce the full response before we declare the ringdown to have begun.

A smaller $\epsilon$ is a stricter requirement which pushes the start time $u_0$ later. 
Since the QNM sum decays exponentially while the tail contribution decays slowly, both amplitudes are smaller and their ratio $\mathcal{Q}$ is also smaller for lower values of $\epsilon$.

\begin{figure*}[!htb]
  \begin{subfigure}[!htb]{\columnwidth}
    \includegraphics[width=\textwidth]{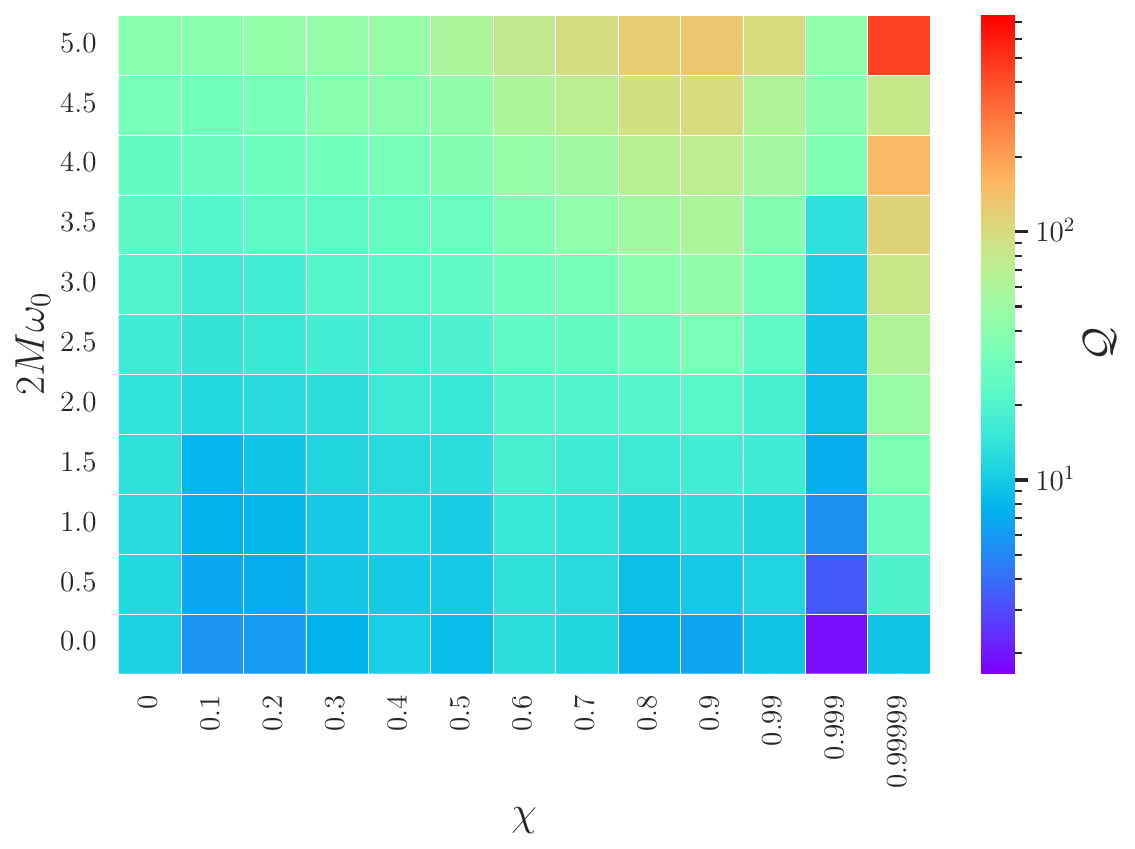}
  \end{subfigure}
  \hfill
  \begin{subfigure}[!htb]{\columnwidth}
    \includegraphics[width=\textwidth]{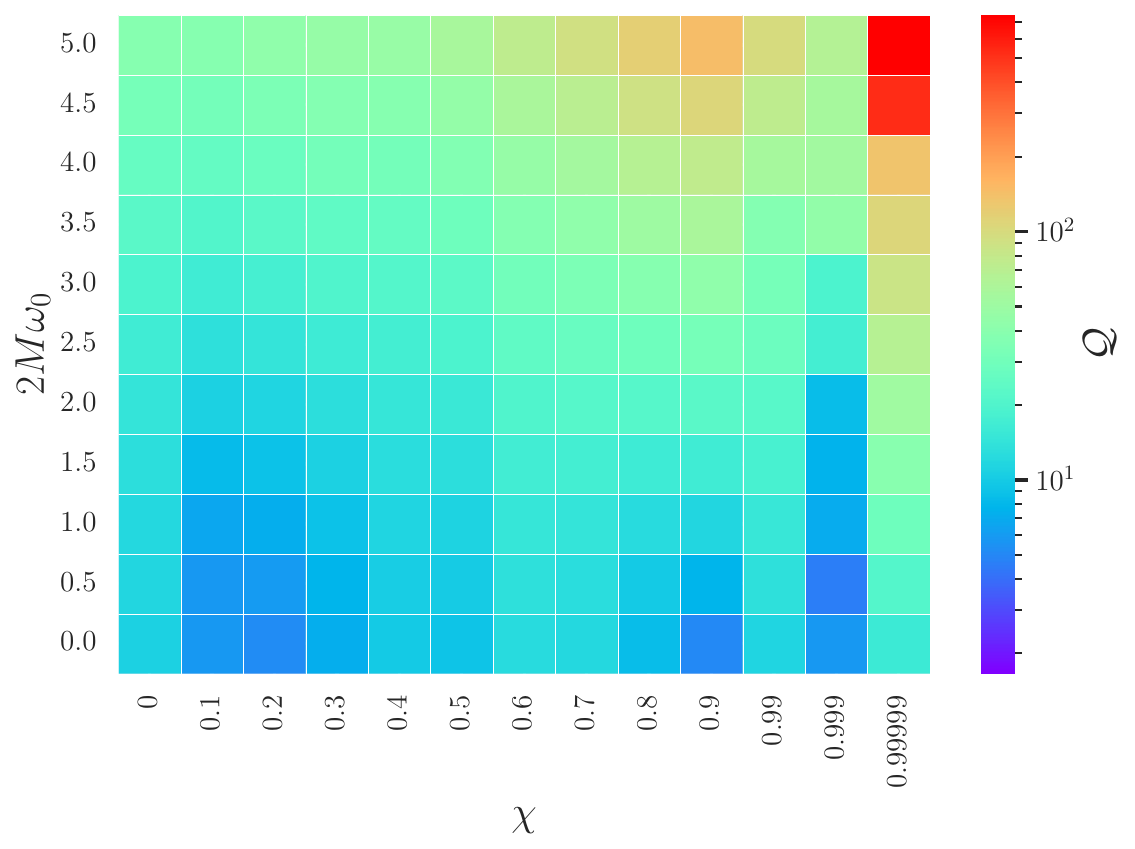}
  \end{subfigure}
  \caption{The ratio $\mathcal{Q}(\chi,\omega_0)$ of QNM to tail amplitude, evaluated at the start time $u_0$, and Gaussian initial data with $\sigma=1, x_0=0$. \emph{Left:} tolerance $\epsilon=0.05$. \emph{Right:} tolerance $\epsilon=0.001$. The qualitative features are the same as in Figure~\ref{fig:ratios}.}
  \label{fig:ratios_tol_compare}
\end{figure*}

\begin{figure*}[!htb]
  \begin{subfigure}[!htb]{\columnwidth}
    \includegraphics[width=\textwidth]{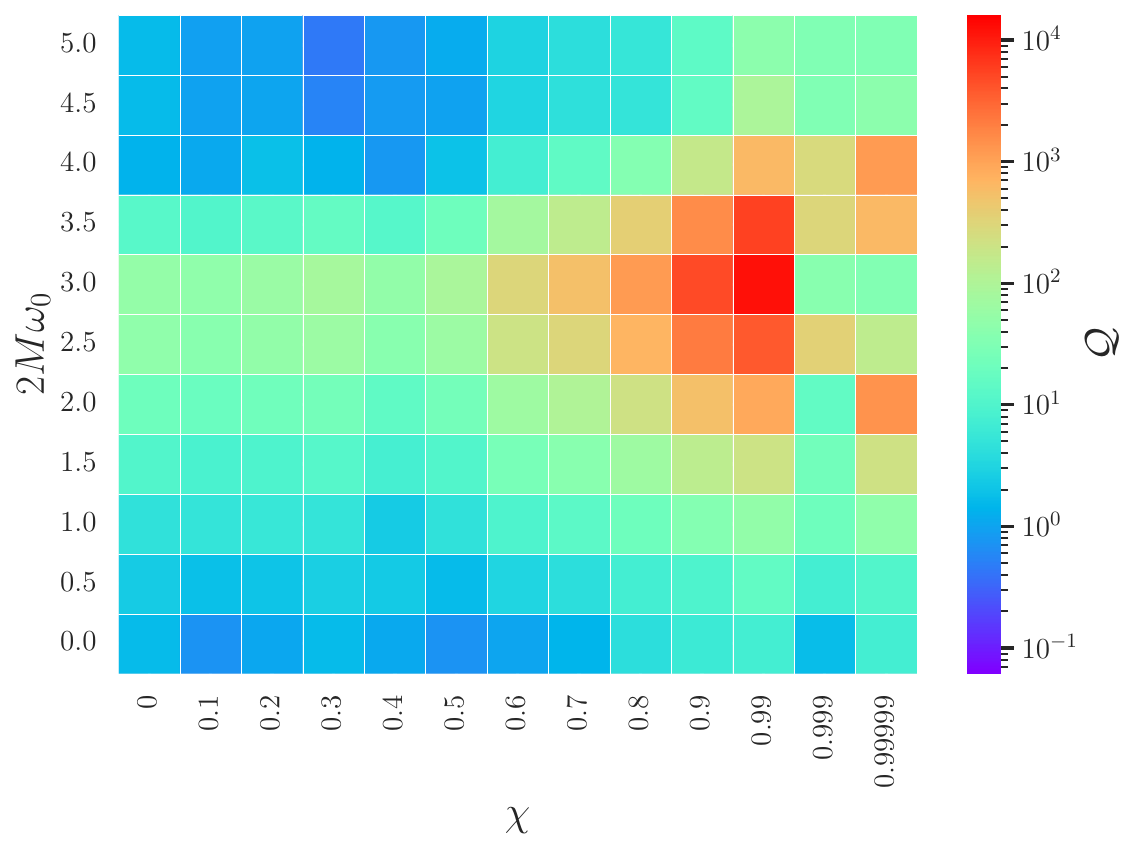}
  \end{subfigure}
  \hfill
  \begin{subfigure}[!htb]{\columnwidth}
    \includegraphics[width=\textwidth]{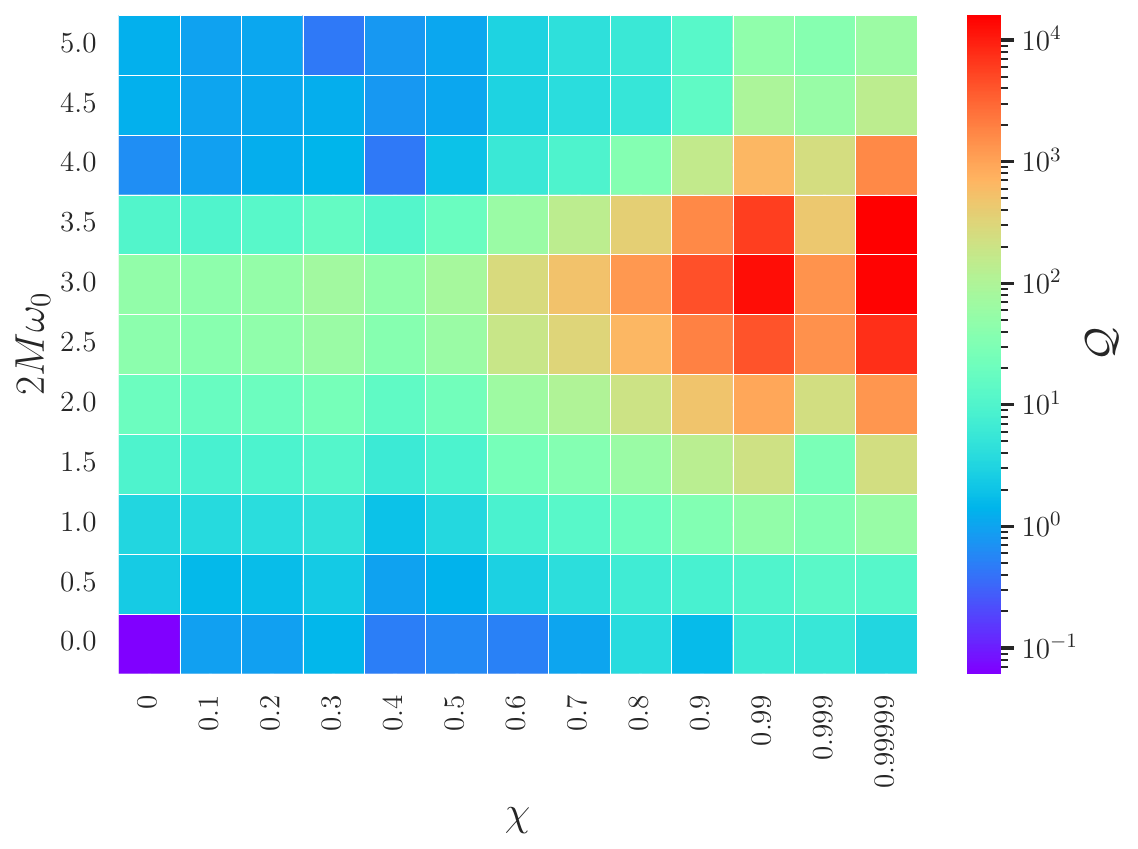}
  \end{subfigure}
  \caption{The ratio $\mathcal{Q}(\chi,\omega_0)$ of QNM to tail amplitude, evaluated at the start time $u_0$, and Gaussian initial data with $\sigma=2, x_0=0$. \emph{Left:} tolerance $\epsilon=0.05$. \emph{Right:} tolerance $\epsilon=0.001$. The qualitative features are the same as in Figure~\ref{fig:ratios_sig_2}.}
  \label{fig:ratios_tol_compare_sig_2}
\end{figure*}

\begin{figure*}[!htb]
  \begin{subfigure}[!htb]{\textwidth}
    \includegraphics[width=\textwidth]{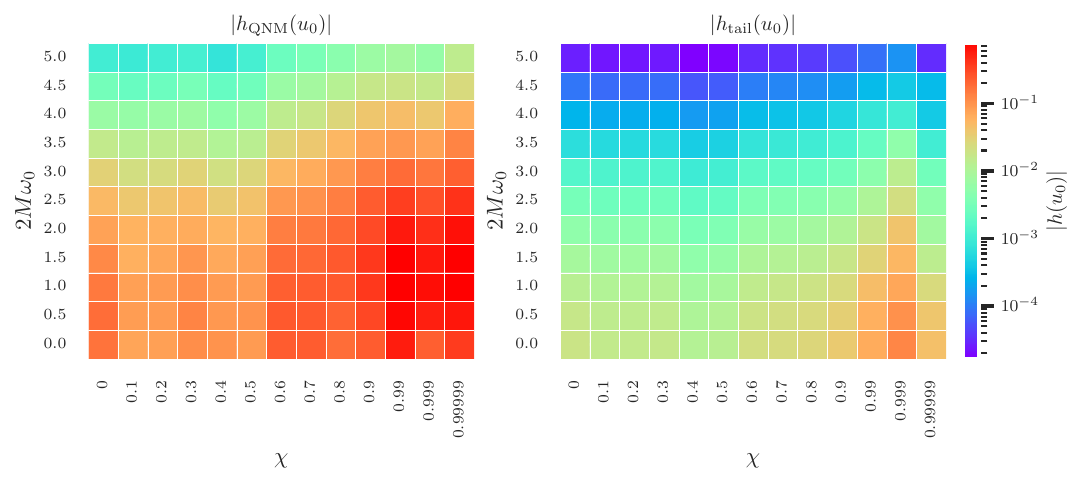}
  \end{subfigure}
  \hfill
  \begin{subfigure}[!htb]{\textwidth}
    \includegraphics[width=\textwidth]{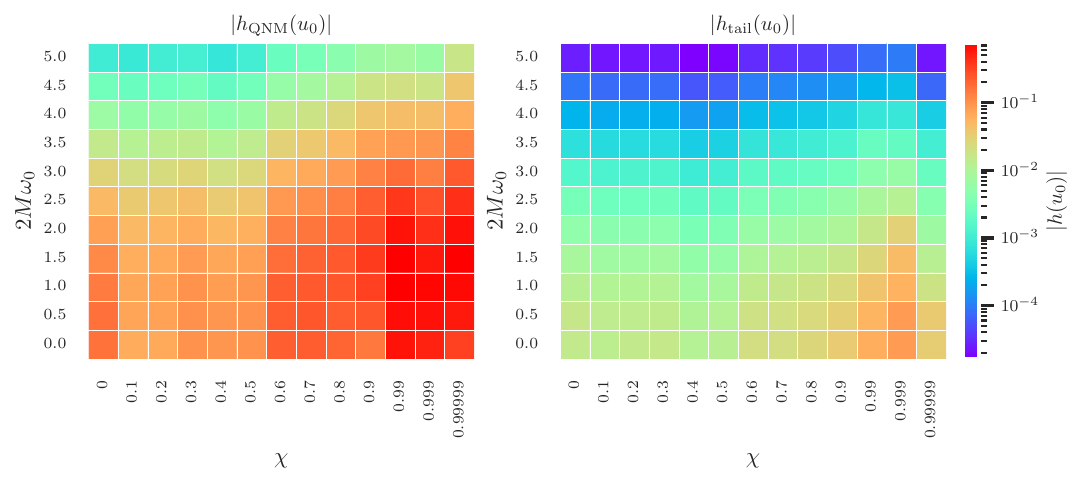}
  \end{subfigure}
  \caption{\emph{Left:} The QNM amplitude at the start time. \emph{Right:} the tail amplitude at the start time. The top two panels use a  tolerance of $\epsilon=0.05$, while the bottom two panels use a tolerance of $\epsilon=0.001$. The qualitative features are the same as in Figure~\ref{fig:amplitudes}.}
  \label{fig:amplitudes_tol_compare}
\end{figure*}

Figures~\ref{fig:ratios_tol_compare}, \ref{fig:ratios_tol_compare_sig_2}, and \ref{fig:amplitudes_tol_compare} repeat the analysis as before but for $\epsilon=0.05,0.001$. Every qualitative feature from the $\epsilon=0.01$ tolerance remains in both cases: for $\sigma=1$, we find that $\mathcal{Q}$ grows with both $\chi$ and $\omega_0$, and the low-frequency turnover near $\chi\simeq0.6-0.7$ persists. In Figure~\ref{fig:ratios_tol_compare_sig_2} we test the robustness of the resonant QNM excitation behaviour with respect to the start time criterion, and similarly find that our results hold for tolerances of $\epsilon=0.05,0.001$. We conclude that our results are stable under (small) perturbations of our choice of start time criterion.

\bibliography{References}

\end{document}